\documentclass[aps,prl,twocolumn,groupedaddress,showpacs]{revtex4}
\usepackage{graphicx}
\def\be{\begin{equation}}
\def\ee{\end{equation}}
\def\ba{\begin{eqnarray}}
\def\ea{\end{eqnarray}}

\def\bc{\begin{center}}
\def\ec{\end{center}}
\begin{document}

\title{Drift plasma instability near the edge as the origin of the microwave-induced zero-resistance states
}

\author{S. A. Mikhailov}
\email[Electronic address: ]{S.Mikhailov@fkf.mpg.de}

\affiliation{Max-Planck Institute for Solid State Research, Heisenbergstr. 1, D-70569 Stuttgart, Germany}

\date{\today}

\begin{abstract}
We discuss a possible origin of the recently discovered microwave-induced 
 zero-resistance states in very-high-electron-mobility two-dimensional electron systems. We suggest a scenario, in which two mechanisms, bulk and edge, contribute to the measured photosignal. Zero-resistance states are assumed to be due to a drift plasma instability, developing {\em near the edge} of the system under the microwave radiation. The proposed scheme qualitatively agrees with the microwave power, temperature, frequency, magnetic field, and mobility dependencies of the measured photosignal.
\end{abstract}


\maketitle

Recently discovered microwave induced zero-resistance states \cite{Mani02,Zudov03} in a two-dimensional (2D) electron system in moderate magnetic fields ($B<0.5$ T) attracted immediate attention of several theoretical groups \cite{Phillips02,Durst03,Andreev03,Anderson03,Shrivastava03,Shi03,Koulakov03,Volkov03} (precursory observations of negative microwave photoresistance have been reported in \cite{Zudov01,Ye01,Mani02proc}). A number of interesting scenarios were suggested, in which the effect is treated in terms of the influence of microwaves on the scattering probability of the electrons in the direction and opposite to the direction of a weak $dc$ field, which results in microwave-stimulated negative local longitudinal conductivity $\sigma_{xx}$ \cite{Durst03,Anderson03,Shi03} (a similar mechanism of the negative photoconductivity was suggested in \cite{Ryzhii70,Ryzhii86}). It was then shown that the negative local $\sigma_{xx}$ should macroscopically manifest itself as the zero-resistance state \cite{Andreev03}. Among other proposed mechanisms is the formation of charge-density waves \cite{Phillips02,Shrivastava03} and the influence of the nonparabolicity of the electronic spectrum \cite{Koulakov03}.

All the theoretical scenarios, proposed so far, imply the {\em bulk} of the sample as the origin of the zero-photoresistance effect. The aim of this Letter is to point to another possible scenario, based on the development of a microwave induced drift instability near the {\em edge} of the sample. We present a simple physical picture, which is in a qualitative agreement with the temperature, microwave power, magnetic field, frequency, and mobility dependencies of the microwave photo-resistance, experimentally observed in Refs. \cite{Mani02,Zudov03}. 

Before starting to discuss the essense of the proposed scenario, we briefly outline the most important and the most puzzling features of the experiments \cite{Mani02,Zudov03}. First, the effect was observed under the conditions
\be
\hbar/\tau\ll kT\simeq \hbar \omega_c\lesssim \hbar\omega \ll E_F,
\label{conditions}
\ee
which imply that it has, probably, a quasiclassical nature. Here, $\tau$ is the momentum relaxation time (estimated from the mobility), $\omega$ and $\omega_c$ are the microwave and the cyclotron frequencies, $T$ is the temperature, and $E_F$ is the Fermi energy. The Coulomb-interaction parameter $r_s=(\pi n_sa_B^2)^{-1/2}$ ($n_s$ is the electron density, $a_B$ is the effective Bohr radius) was about 1 in the experiment, suggesting that electron-electron correlations should not play a decisive role in the effect. 
Second, the effect was seen in the longitudinal resistance $R_{xx}$ and no indications on the influence of microwaves on the Hall resistance was observed. 
Third, and the most intriguing feature of the new experiments \cite{Mani02,Zudov03} is that their results are {\em in a strong contradiction} with another, very similar experiment \cite{Vasiliadou93}, performed about 10 years ago. In Ref. \cite{Vasiliadou93}, the microwave photoresponse of Hall bars was studied under conditions, very similar to those of the new experiments. The density of electrons, the range of frequencies and magnetic fields, as well as the size of the samples in the new and the old experiments are very close. The only difference is the mobility, which is by more than one order of magnitude higher in the new samples. The photoresistance signal in the old samples \cite{Vasiliadou93} had the form of a {\em weak peak} centered at the {\em magnetoplasmon} frequency $\omega_{mp}=(\omega_p^2+\omega_c^2)^{1/2}$, where the plasma frequency $\omega_p\propto w^{-1/2}$ was determined by the width of the sample $w$ and lay in the 60--100 GHz range. This is in a very good agreement with an intuitive physical picture of the variation of the photoresponse due to a heating of electrons at the absorption frequency. Contrary to that, in the new experiments {\em huge oscillations}, governed exclusively by the {\em cyclotron frequency} were observed. Although the estimated magnetoplasmon frequency in the new samples lies in the same frequency range, no indications of the magnetoplasmon resonance were seen in \cite{Mani02,Zudov03}. Therefore, a theory should, ideally, explain, why the magnetoplasmon shift of the cyclotron frequency was seen in the old, relatively ``dirty'' samples, and why it is not seen in the new, extremely-high-quality samples. Theoretical scenarios proposed so far \cite{Phillips02,Durst03,Andreev03,Anderson03,Shrivastava03,Shi03,Koulakov03,Ryzhii70,Ryzhii86} assumed that the sample is infinite and did not address this puzzle at all. 

The dramatic difference between results of the old and new experiments suggests that a principally different mechanism should be responsible for the impressive features discovered in Refs. \cite{Mani02,Zudov03}. Therefore, we propose a scheme, in which two mechanisms contribute to the experimentally measured voltage between longitudinal contacts. The first contribution has a bulk origin and was seen in the old samples. The second contribution originates at the edge of the sample. We show that microwaves may induce a drift plasma instability near the edge of the sample. Intervals of frequencies and magnetic fields, in which the instability exists, coincide with those, where the zero-resistance states and minima in $R_{xx}$ were observed. The instability may thus be responsible for the zero-resistance states. Its existence  requires low scattering rates, and therefore was not seen in the old, relatively ``dirty'' samples. 

First, we consider the bulk effect. Under the conditions (\ref{conditions}) the response of an infinite 2D electron gas on the microwave radiation can be described by the classical Boltzmann equation neglecting the scattering integral. The solution for the electron distribution function, valid for an arbitrarily strong microwave field, is $f_\approx({\bf p},t) = f_0({\bf p}-m{\bf V}(t))$, where $f_0$ is the Fermi-Dirac distribution function, the velocity 
\be
V_x(t)+iV_y(t)=\frac{-i(e/m)E_\approx^{ext}}{\omega-\omega_c+i\Gamma}e^{i\omega t}
\label{Vxy}
\ee
is the solution of classical equations of motion for one electron, $E_\approx^{ext}$ is the amplitude of the incident electromagnetic wave, and $\Gamma=2\pi n_se^2/mc\sqrt{\kappa}$ is the radiative decay \cite{Mikhailov96a}, which gives the dominant contribution to the cyclotron resonance linewidth in very-high-electron-mobility samples, $\Gamma\gg 1/\tau$. Here $n_s$ and $m$ are the density and the effective mass of 2D electrons, $c$ is the velocity of light, and $\kappa$ is the dielectric constant of the surrounding medium. In (\ref{Vxy}) we have assumed that the incident wave is circularly polarised. 

\begin{figure}
\includegraphics[width=8.2cm]{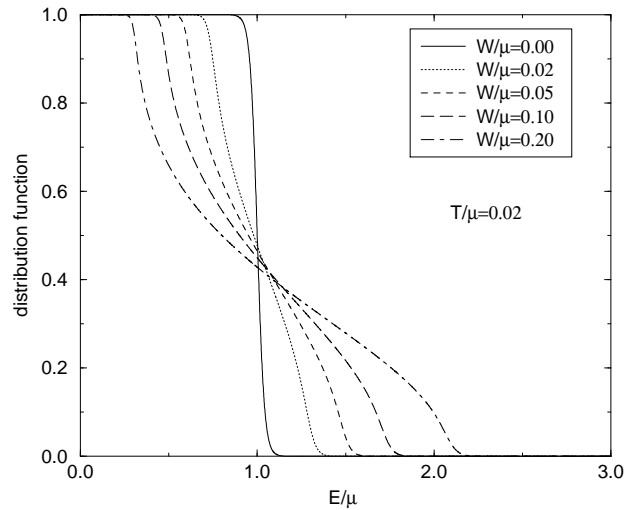}
\caption{Electron distribution function (\protect\ref{fdav}) at different microwave power levels.}
\label{fd}
\end{figure}

Averaging the function $f_\approx({\bf p},t)$ over the period of the microwave field, one gets the time-independent, microwave-modified electron distribution function 
\be
F_0(E,W)=\frac 1\pi\int_0^\pi\frac {d x}{1+\exp\left[\frac{E -\mu+W+2\sqrt{WE}\cos  x}{kT}\right]},
\label{fdav}
\ee
shown in Figure \ref{fd}. The function $F_0$ can be used for estimating the microwave-induced corrections to the $dc$ conductivity $\sigma_{xx}$, which appear due to the energy dependence of the momentum-relaxation time $\tau(E)$ \cite{note1} (the Hall conductivity $\sigma_{xy}$ is not influenced by microwaves at $\omega_c\tau\gg 1$). The microwave-related energy parameter 
\be
W=\frac{e^2|E_\approx^{ext}|^2}{2m[(\omega-\omega_c)^2+\Gamma^2]}
\label{W}
\ee
is proportional to the microwave power and has the form of a peak centered at the cyclotron frequency. Taking into account finite dimensions of the sample [the field $E_\approx^{ext}$ should be screened by the dielectric function of a finite-size sample, $E_\approx^{ext}\to E_\approx^{ext}/\epsilon(q,\omega)$, $q\simeq \pi/w$] shifts the resonance position to the magnetoplasmon frequency. Estimates of the magnitude of the microwave electric field shows that the energy $W$ is always much smaller than the relevant bulk energy scale $E_F$, $W\ll E_F$. These qualitative arguments show that the bulk contribution to the photoconductivity should have the form of a {\em weak peak} at the {\em magnetoplasmon} frequency $\omega_{mp}$. This is in obvious agreement with intuitive expectations, as well as with the results of the old experiment \cite{Vasiliadou93}. 

\begin{figure}
\includegraphics[width=8.2cm]{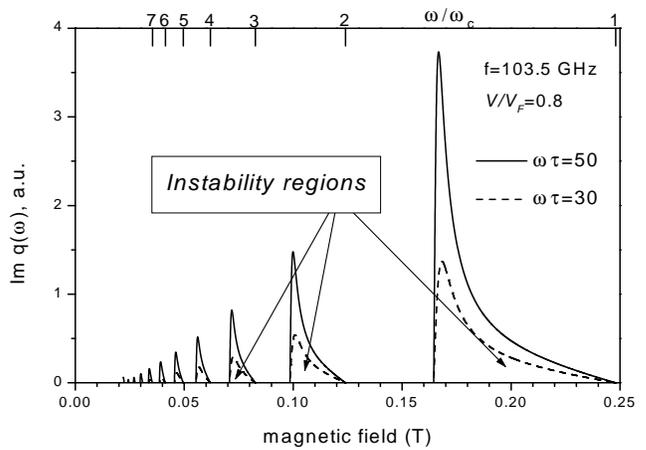}
\caption{Growth rate of the instability as a function of magnetic field for parameters of Ref. \cite{Mani02}, the velocity $V/V_F=0.8$, and two different scattering rates. Numbers on the top axis show the position of the cyclotron harmonics.}
\label{mani}
\end{figure}

Now, consider what happens near the edge of the system under the microwave radiation. The same estimates show that the energy $W$ can be comparable with or even substantially larger than the temperature,
\be
W\gg kT.
\ee
Hence, the microwave radiation can significantly change the distribution of electrons over the quantum states near the edge of the system. Accepting the standard picture of Landau levels bent up near the edge and crossing the Fermi level, we see that microwaves lead to the appearence of electrons near the edge which occupy higher energy levels and hence are running along the boundary (skipping orbits) with an increased (compared to the dark situation) velocity. Thus we have a situation, typical for the development of drift plasma instabilities. In the bulk of the system we have a 2D plasma, characterized by the dielectric function $\epsilon(q,\omega)$, and near the edge an ``electron beam'', which can move with respect to the bulk, with a sufficiently large velocity $V$ induced by the microwave radiation. The spectrum of plasma waves in this situation is described by the dispersion equation of the type \cite{Landau10}
\be
\omega-qV=\pm\frac{\omega'_p}{\sqrt{\epsilon(q,\omega)}},
\label{de}
\ee
and under certain conditions, namely if $\epsilon(q,\omega)<0$, may have unstable solutions (here $\omega'_p$ is the plasma frequency in the beam). The dielectric function of a 2D system in a magnetic field has the form $\epsilon(q,\omega)=1+2\pi i\sigma_{xx}(q,\omega)q/\omega\kappa$, where the wave-vector and frequency dependent conductivity is \cite{Chiu74}
\be
\sigma_{xx}(q,\omega)=\frac{n_se^2}{m\omega_c}\frac{\omega+i\gamma}{i\omega_c}\left(\frac 2{qr_c}\right)^2 \sum_{k=1}^\infty\frac{k^2J_k^2(qr_c)}{k^2-[(\omega+i\gamma)/\omega_c]^2},
\label{sigma}
\ee
and $r_c$ is the cyclotron radius. As seen from here, the dielectric function is negative in certain frequency intervals {\em above} multiples of the cyclotron frequency (instability regions). Resolving Eq. (\ref{de}) with respect to $q$ at fixed (real) frequency $\omega$ and at $\omega'_p\ll\omega_p$, we get the imaginary part of the wave vector $q''(\omega)$, which characterizes the growth rate of the instability. Figure \ref{mani} shows $q''(\omega)$, as a function of magnetic field $B$, for parameters typical for the experiment \cite{Mani02}, and for a reasonable value of the drift velocity $V$ of 0.8 times the Fermi velocity $V_F$. The areas, where the growth rate $q''(\omega)$ is positive, correspond to the instability regions. Comparing Figure \ref{mani} with Figure 3a from Ref. \cite{Mani02} one sees that these instability regions very well correspond to the intervals of $B$ where the zero-resistance states (and the negative photoresistance at higher cyclotron harmonics) were experimentally observed (a similar picture plotted for Figure 1 from Ref. \cite{Zudov03} gave the same result). A remarkable feature of the instability plot of Figure \ref{mani} is that the ``strength'' of the instabilities decreases rather slowly with the harmonics number, in a qualitative agreement with experiment. Besides, two different curves in Figure \ref{mani}, drawn for different scattering rates, show that the higher electron mobility favours the development of the instability, which also agrees with \cite{Mani02,Zudov03}. Finally, we notice that the number of electrons, running along the edge with the microwave-increased velocity, is proportional to $\exp(-W/T)$, which qualitatively agrees with the power and temperature dependencies of the photosignal at the minima of $R_{xx}$.
 
Thus, the microwave power, temperature, magnetic field, frequency and scattering rate dependencies of the growth rate of the instability qualitatively agree with such dependencies of the experimentally measured photoresistance $R_{xx}$. A detailed mechanism of how the instability is related with $R_{xx}$ is still to be understood, but it is clear that under the unstable conditions the distribution of electric current and potential in the sample will be very complicated, and the instability may macroscopically manifest itself via a vanishing potential difference between certain pairs of contacts. Ideas along these lines have been recently discussed in Ref. \cite{Andreev03}. 
Although a more comprehensive analysis is certainly required to support the ideas of this work, we believe that the presented arguments catch the main physics of the phenomena, in particular, the fact that the instability exists in the intervals where $\epsilon(q,\omega)<0$. Instabilities of the considered type are well known in a three-dimensional gaseous plasma, see e.g. \cite{Mikhailovskii92,Horton99}. The effects observed in Refs. \cite{Mani02,Zudov03} is probably a manifestation of such an instability in the solid-state two-dimensional plasma.

The $j\pm 1/4$ law for the positions of the photoresistance minima/maxima \cite{Mani02} can be qualitatively understood if to assume that the instability regions occupy approximately one half of the interval between $\omega=j\omega_c$ and $\omega=(j+1)\omega_c$ (see Figure \ref{mani}). Then, in the regime of weak instability, when the photoresponse has a sinusoidal form, it should have the observed phase. On the other hand, if the instability regions occupy less or more than one half of the interval $[j\omega_c , (j+1)\omega_c]$ (this may depend on parameters of the system), the photoresistance minima/maxima should not necessarily be at the $\omega/\omega_c=j\pm 1/4$ positions. The $j\pm 1/4$ observation \cite{Mani02}, as follows from our approach, is thus not necessarily a universal law. Notice that the $R_{xx}$ maxima were seen at $\omega/\omega_c=j- 1/4$, according to \cite{Mani02}, and at $\omega/\omega_c=j$, according to \cite{Zudov03}.

The proposed scenario also allows us to resolve the paradox related with the non-manifestation of the magnetoplasmon resonance with $q\simeq 1/w$ in the new experiments: while in \cite{Vasiliadou93} the 2D plasmon wave vector was determined by the width of the sample $w$, in the new experiments the relevant wavevectors are fixed by the instability conditions (\ref{de}), and the weak ``bulk'' magnetoplasmon resonance is not seen on the background of huge oscillations due to the edge contribution. Notice, that the weak bulk-magnetoplasmon resonance which was seen in \cite{Zudov01} {\em along with} relatively weak sinusoidal oscillations of $R_{xx}$ (in moderate-mobility samples), supports our idea of two, bulk and edge, contributions to the measured photosignal. 

To summarize, we have shown that the recently observed microwave-induced zero-resistance states and effects of the negative photoresistance in very-high-electron mobility 2D electron systems in strong ($\omega_c\tau\gg 1$) magnetic fields can be related to the development of a drift plasma instability, arising near the edge of the system under the microwave radiation. This interpretation also allowed us to consistently explain the dramatic difference between results of the old \cite{Vasiliadou93} and new \cite{Mani02,Zudov03} experiments, which were performed under the seemingly similar conditions. Our results are in a qualitative agreement with the microwave power, temperature, magnetic field, frequency, and electron-mobility dependencies observed in the experiments.

I would like to thank Ramesh Mani, Jurgen Smet, and Klaus von Klitzing for an opportunity to see experimental data prior to publication, and for numerous discussions of details of the experiment \cite{Mani02}. I am especially thankful to Klaus von Klitzing for critical discussions of different versions of the theory. I also thank Igor Kukushkin and Sergey Dorozhkin for very useful discussions of details of other microwave experiments, as well as Rolf Gerhardts for reading the manuscript and useful comments.

\bibliography{../../../BIB-FILES/lowD,../../../BIB-FILES/emp,../../../BIB-FILES/mikhailov}

\begin{thebibliography}{22}
\expandafter\ifx\csname natexlab\endcsname\relax\def\natexlab#1{#1}\fi
\expandafter\ifx\csname bibnamefont\endcsname\relax
  \def\bibnamefont#1{#1}\fi
\expandafter\ifx\csname bibfnamefont\endcsname\relax
  \def\bibfnamefont#1{#1}\fi
\expandafter\ifx\csname citenamefont\endcsname\relax
  \def\citenamefont#1{#1}\fi
\expandafter\ifx\csname url\endcsname\relax
  \def\url#1{\texttt{#1}}\fi
\expandafter\ifx\csname urlprefix\endcsname\relax\def\urlprefix{URL }\fi
\providecommand{\bibinfo}[2]{#2}
\providecommand{\eprint}[2][]{\url{#2}}

\bibitem[{\citenamefont{Mani et~al.}(2002)\citenamefont{Mani, Smet, von
  Klitzing, Narayanamurti, Johnson, and Umansky}}]{Mani02}
\bibinfo{author}{\bibfnamefont{R.~G.} \bibnamefont{Mani}},
  \bibinfo{author}{\bibfnamefont{J.~H.} \bibnamefont{Smet}},
  \bibinfo{author}{\bibfnamefont{K.}~\bibnamefont{von Klitzing}},
  \bibinfo{author}{\bibfnamefont{V.}~\bibnamefont{Narayanamurti}},
  \bibinfo{author}{\bibfnamefont{W.~B.} \bibnamefont{Johnson}},
  \bibnamefont{and} \bibinfo{author}{\bibfnamefont{V.}~\bibnamefont{Umansky}},
  \bibinfo{journal}{Nature} \textbf{\bibinfo{volume}{420}},
  \bibinfo{pages}{646} (\bibinfo{year}{2002}).

\bibitem[{\citenamefont{Zudov et~al.}(2003)\citenamefont{Zudov, Du, Pfeiffer,
  and West}}]{Zudov03}
\bibinfo{author}{\bibfnamefont{M.~A.} \bibnamefont{Zudov}},
  \bibinfo{author}{\bibfnamefont{R.~R.} \bibnamefont{Du}},
  \bibinfo{author}{\bibfnamefont{L.~N.} \bibnamefont{Pfeiffer}},
  \bibnamefont{and} \bibinfo{author}{\bibfnamefont{K.~W.} \bibnamefont{West}},
  \bibinfo{journal}{Phys. Rev. Lett.} \textbf{\bibinfo{volume}{90}},
  \bibinfo{pages}{046807} (\bibinfo{year}{2003}).

\bibitem[{\citenamefont{Phillips}()}]{Phillips02}
\bibinfo{author}{\bibfnamefont{J.~C.} \bibnamefont{Phillips}},
  \bibinfo{note}{cond-mat/0212416}.

\bibitem[{\citenamefont{Durst et~al.}()\citenamefont{Durst, Sachdev, Read, and
  Girvin}}]{Durst03}
\bibinfo{author}{\bibfnamefont{A.~C.} \bibnamefont{Durst}},
  \bibinfo{author}{\bibfnamefont{S.}~\bibnamefont{Sachdev}},
  \bibinfo{author}{\bibfnamefont{N.}~\bibnamefont{Read}}, \bibnamefont{and}
  \bibinfo{author}{\bibfnamefont{S.~M.} \bibnamefont{Girvin}},
  \bibinfo{note}{cond-mat/0301569}.

\bibitem[{\citenamefont{Andreev et~al.}()\citenamefont{Andreev, Aleiner, and
  Millis}}]{Andreev03}
\bibinfo{author}{\bibfnamefont{A.~V.} \bibnamefont{Andreev}},
  \bibinfo{author}{\bibfnamefont{I.~L.} \bibnamefont{Aleiner}},
  \bibnamefont{and} \bibinfo{author}{\bibfnamefont{A.~J.}
  \bibnamefont{Millis}}, \bibinfo{note}{cond-mat/0302063}.

\bibitem[{\citenamefont{Anderson and Brinkman}()}]{Anderson03}
\bibinfo{author}{\bibfnamefont{P.~W.} \bibnamefont{Anderson}} \bibnamefont{and}
  \bibinfo{author}{\bibfnamefont{W.~F.} \bibnamefont{Brinkman}},
  \bibinfo{note}{cond-mat/0302129}.

\bibitem[{\citenamefont{Shrivastava}()}]{Shrivastava03}
\bibinfo{author}{\bibfnamefont{K.~N.} \bibnamefont{Shrivastava}},
  \bibinfo{note}{cond-mat/0302320}.

\bibitem[{\citenamefont{Shi and Xie}()}]{Shi03}
\bibinfo{author}{\bibfnamefont{J.}~\bibnamefont{Shi}} \bibnamefont{and}
  \bibinfo{author}{\bibfnamefont{X.}~\bibnamefont{Xie}},
  \bibinfo{note}{cond-mat/0302393}.

\bibitem[{\citenamefont{Koulakov and Raikh}()}]{Koulakov03}
\bibinfo{author}{\bibfnamefont{A.~A.} \bibnamefont{Koulakov}} \bibnamefont{and}
  \bibinfo{author}{\bibfnamefont{M.~E.} \bibnamefont{Raikh}},
  \bibinfo{note}{cond-mat/0302465}.

\bibitem[{\citenamefont{Volkov}()}]{Volkov03}
\bibinfo{author}{\bibfnamefont{A.~F.} \bibnamefont{Volkov}},
  \bibinfo{note}{cond-mat/0302615}.

\bibitem[{\citenamefont{Zudov et~al.}(2001)\citenamefont{Zudov, Du, Simmons,
  and Reno}}]{Zudov01}
\bibinfo{author}{\bibfnamefont{M.~A.} \bibnamefont{Zudov}},
  \bibinfo{author}{\bibfnamefont{R.~R.} \bibnamefont{Du}},
  \bibinfo{author}{\bibfnamefont{J.~A.} \bibnamefont{Simmons}},
  \bibnamefont{and} \bibinfo{author}{\bibfnamefont{J.~L.} \bibnamefont{Reno}},
  \bibinfo{journal}{Phys. Rev. B} \textbf{\bibinfo{volume}{64}},
  \bibinfo{pages}{201311} (\bibinfo{year}{2001}).

\bibitem[{\citenamefont{Ye et~al.}(2001)\citenamefont{Ye, Engel, Tsui, Simmons,
  Wendt, Vawter, and Reno}}]{Ye01}
\bibinfo{author}{\bibfnamefont{P.~D.} \bibnamefont{Ye}},
  \bibinfo{author}{\bibfnamefont{L.~W.} \bibnamefont{Engel}},
  \bibinfo{author}{\bibfnamefont{D.~C.} \bibnamefont{Tsui}},
  \bibinfo{author}{\bibfnamefont{J.~A.} \bibnamefont{Simmons}},
  \bibinfo{author}{\bibfnamefont{J.~R.} \bibnamefont{Wendt}},
  \bibinfo{author}{\bibfnamefont{G.~A.} \bibnamefont{Vawter}},
  \bibnamefont{and} \bibinfo{author}{\bibfnamefont{J.~L.} \bibnamefont{Reno}},
  \bibinfo{journal}{Appl. Phys. Lett.} \textbf{\bibinfo{volume}{79}},
  \bibinfo{pages}{2193} (\bibinfo{year}{2001}).

\bibitem[{\citenamefont{Mani et~al.}(29 July - 2 August
  2002)\citenamefont{Mani, Smet, von Klitzing, Narayanamurti, Johnson, and
  Umansky}}]{Mani02proc}
\bibinfo{author}{\bibfnamefont{R.~G.} \bibnamefont{Mani}},
  \bibinfo{author}{\bibfnamefont{J.~H.} \bibnamefont{Smet}},
  \bibinfo{author}{\bibfnamefont{K.}~\bibnamefont{von Klitzing}},
  \bibinfo{author}{\bibfnamefont{V.}~\bibnamefont{Narayanamurti}},
  \bibinfo{author}{\bibfnamefont{W.~B.} \bibnamefont{Johnson}},
  \bibnamefont{and} \bibinfo{author}{\bibfnamefont{V.}~\bibnamefont{Umansky}},
  in \emph{\bibinfo{booktitle}{Proc. of the 26th Int. Conf. on the Physics of
  Semiconductors}} (\bibinfo{address}{Edingurgh}, \bibinfo{year}{29 July - 2
  August 2002}), \bibinfo{note}{to be published}.

\bibitem[{\citenamefont{Ryzhii}(1970)}]{Ryzhii70}
\bibinfo{author}{\bibfnamefont{V.~I.} \bibnamefont{Ryzhii}},
  \bibinfo{journal}{Sov. Phys. -- Solid State} \textbf{\bibinfo{volume}{11}},
  \bibinfo{pages}{2078} (\bibinfo{year}{1970}).

\bibitem[{\citenamefont{Ryzhii et~al.}(1986)\citenamefont{Ryzhii, Suris, and
  Shchamkhalova}}]{Ryzhii86}
\bibinfo{author}{\bibfnamefont{V.~I.} \bibnamefont{Ryzhii}},
  \bibinfo{author}{\bibfnamefont{R.~A.} \bibnamefont{Suris}}, \bibnamefont{and}
  \bibinfo{author}{\bibfnamefont{B.~S.} \bibnamefont{Shchamkhalova}},
  \bibinfo{journal}{Sov. Phys. Semicond.} \textbf{\bibinfo{volume}{20}},
  \bibinfo{pages}{1299} (\bibinfo{year}{1986}).

\bibitem[{\citenamefont{Vasiliadou et~al.}(1993)\citenamefont{Vasiliadou,
  M\"uller, Heitmann, Weiss, von Klitzing, Nickel, Schlapp, and
  L\"osch}}]{Vasiliadou93}
\bibinfo{author}{\bibfnamefont{E.}~\bibnamefont{Vasiliadou}},
  \bibinfo{author}{\bibfnamefont{G.}~\bibnamefont{M\"uller}},
  \bibinfo{author}{\bibfnamefont{D.}~\bibnamefont{Heitmann}},
  \bibinfo{author}{\bibfnamefont{D.}~\bibnamefont{Weiss}},
  \bibinfo{author}{\bibfnamefont{K.}~\bibnamefont{von Klitzing}},
  \bibinfo{author}{\bibfnamefont{H.}~\bibnamefont{Nickel}},
  \bibinfo{author}{\bibfnamefont{W.}~\bibnamefont{Schlapp}}, \bibnamefont{and}
  \bibinfo{author}{\bibfnamefont{R.}~\bibnamefont{L\"osch}},
  \bibinfo{journal}{Phys. Rev. B} \textbf{\bibinfo{volume}{48}},
  \bibinfo{pages}{17145} (\bibinfo{year}{1993}).

\bibitem[{\citenamefont{Mikhailov}(1996)}]{Mikhailov96a}
\bibinfo{author}{\bibfnamefont{S.~A.} \bibnamefont{Mikhailov}},
  \bibinfo{journal}{Phys. Rev. B} \textbf{\bibinfo{volume}{54}},
  \bibinfo{pages}{10335} (\bibinfo{year}{1996}).

\bibitem[{not()}]{note1}
\bibinfo{note}{We do not consider the influence of microwaves on the scattering
  rate, as it was done in \cite{Durst03,Anderson03,Ryzhii70,Ryzhii86}.
  Scenarios, developed in these papers, require either sufficiently strong $dc$
  electric fields ($eE_{dc}l_B\gg\hbar/\tau$, where $l_B$ is the magnetic
  length) \cite{Ryzhii70,Ryzhii86}, or sufficiently strong disorder
  \cite{Durst03,Anderson03}. Since the experiments were performed in the
  highest-mobility samples in weak $dc$ electric fields
  ($eE_{dc}l_B\ll\hbar/\tau$), we try to find another explanation of the
  observed effects.}

\bibitem[{\citenamefont{Lifshitz and Pitaevskii}(1981)}]{Landau10}
\bibinfo{author}{\bibfnamefont{E.~M.} \bibnamefont{Lifshitz}} \bibnamefont{and}
  \bibinfo{author}{\bibfnamefont{L.~P.} \bibnamefont{Pitaevskii}},
  \emph{\bibinfo{title}{Physical Kinetics}} (\bibinfo{publisher}{Pergamon
  Press}, \bibinfo{year}{1981}), \bibinfo{note}{\S 61}.

\bibitem[{\citenamefont{Chiu and Quinn}(1974)}]{Chiu74}
\bibinfo{author}{\bibfnamefont{K.~W.} \bibnamefont{Chiu}} \bibnamefont{and}
  \bibinfo{author}{\bibfnamefont{J.~J.} \bibnamefont{Quinn}},
  \bibinfo{journal}{Phys. Rev. B} \textbf{\bibinfo{volume}{9}},
  \bibinfo{pages}{4724} (\bibinfo{year}{1974}).

\bibitem[{\citenamefont{Mikhailovskii}(1992)}]{Mikhailovskii92}
\bibinfo{author}{\bibfnamefont{A.~B.} \bibnamefont{Mikhailovskii}},
  \emph{\bibinfo{title}{Electromagnetic instabilities in an inhomogeneous
  plasma}} (\bibinfo{publisher}{Institute of Physics Publishing},
  \bibinfo{year}{1992}).

\bibitem[{\citenamefont{Horton}(1999)}]{Horton99}
\bibinfo{author}{\bibfnamefont{W.}~\bibnamefont{Horton}},
  \bibinfo{journal}{Rev. Mod. Phys.} \textbf{\bibinfo{volume}{71}},
  \bibinfo{pages}{735} (\bibinfo{year}{1999}).

\end{thebibliography}

\end{document}